# Field-free spin orbit torque switching of synthetic antiferromagnet through interlayer Dzyaloshinskii-Moriya interaction

Zilu Wang[a,b,*], Pingzhi Li[b], Yuxuan Yao[a], Youri L.W. Van Hees[b], Casper F. Schippers[b], Reinoud Lavrijsen[b], Albert Fert[a,c], Weisheng Zhao[a,*], and Bert Koopmans[b]

[a]Fert Beijing Institute, MIIT Key Laboratory of Spintronics, School of Integrated Circuit Science and Engineer Beihang University, Beijing 100191, China;

[b]Department of Applied Physics, Eindhoven University of Technology, P.O. Box 513, 5600 MB Eindhoven, The Netherland;

[c]Unité Mixte de Physique, CNRS, Thales, University of Paris-Saclay, Palaiseau, France.

*Zilu Wang, Weisheng Zhao.

**Email:** z.wang1@tue.nl; weisheng.zhao@buaa.edu.cn

**This PDF file includes:**

>    Main Text
>
>    Figures 1 to 5

**Abstract**

Perpendicular synthetic antiferromagnets (SAFs) are of interest for the next generation ultrafast, high density spintronic memory and logic devices. However, to energy efficiently operate their magnetic order by current-induced spin orbit torques (SOTs), an unfavored high external field is conventionally required to break the symmetry. Here, we theoretically and experimentally demonstrate the field-free SOT switching of a perpendicular SAF through the introduction of interlayer Dzyaloshinskii–Moriya interaction (DMI). By macro-spin simulation, we show that the speed of field-free switching increases with the in-plane mirror asymmetry of injected spins. We experimentally observe the existence of interlayer DMI in our SAF sample by an azimuthal angular dependent anomalous Hall measurement. Field-free switching is accomplished in such a sample and the strength of the effective switching field demonstrates its origin from interlayer DMI. Our results provide a new strategy for SAF based high performance SOT devices.

**Significance Statement**

In the post-Moore era, spintronic is one of the key technologies to break the bottleneck of high energy consumption of electronic devices. However, as the process node moves on, conventional ferromagnet-based spintronic devices meet the superparamagnetic size limit. Under such circumstance, perpendicular synthetic antiferromagnets with high scalability, strong immunity to external field and ultrafast dynamics, become a competitive choice for the next generation spintronics memory and logic devices. Here, we propose a strategy to energy efficiently operate their magnetic order by spin orbit torque (SOT) under zero external field. Compared with previous field-free switching solutions, the performance of SOT device is well preserved. Our work pave the way for a wide range of potential high performance SOT devices.

**Main Text**

**Introduction**

Over the past decade, manipulation of perpendicular magnetization by spin orbit torque (SOT) has been laying the foundation for high density, energy efficient and ultrafast spintronic memory and logic devices (1–5). However, as the devices scale down, higher thermal stability is required to ensure enough data retention time (6, 7). Belonging to the family of antiferromagnets, synthetic antiferromagnets (SAFs) possess high thermal stability (8), low stray field, high immunity to external field (9) and fast dynamics (10, 11), yet their magnetic state is almost as easily detectable as for a ferromagnet (FM) (11). These features make SAF a competitive choice for next generation ultrafast, high density spintronic devices (12).

To deterministically operate the magnetic order of SAFs with perpendicular magnetic anisotropy (p-SAFs) by SOT, an unfavored high external field is needed to break the in-plane symmetry (13–16). This will not only decrease the data density and energy efficiency, but also add extra uncertainty to the switching behavior of the p-SAF (17). To eliminate this field, some insights could be learned from the intensive studies on how to break the symmetry alternatively in a FM single layer. These efforts introduced either an external stray (or exchange) field (18–20), or an internal structural (or magnetic) asymmetry (21–27). However, to date, these alternatives are usually accompanied by a sacrifice of data density (18), energy efficiency (19–24) or large-scale uniformity (25, 27, 28).

Recently, a novel asymmetric exchange interaction called interlayer Dzyaloshinskii-Moriya interaction (DMI) has been demonstrated in a SAF (29–31). This is a long-range interaction that will cause a chiral spin texture between two magnetic layers across an insertion layer. Comparable to an interfacial DMI, an interlayer DMI is also derived from the combination of spin-orbit coupling and structural asymmetry. The required structural asymmetry could be introduced by asymmetric deposition of insertion layers (29, 30). We show that this interaction turns out to enable an improved solution for field-free SOT switching of a p-SAF. In this field-free solution, no restriction is put on the structure and material of the magnet and the SOT channel. Therefore, the performance of SOT devices would be well preserved. Besides, no extra functional layer is attached, which limits the current shunting and makes it compatible with most potential spintronic applications (32, 33).

In this article, we theoretically demonstrate and experimentally observe the field-free SOT switching of a Pt/CoB/Pt/Ru/Pt/CoB/Pt p-SAF by means of an interlayer DMI. Tilted sputter growth of a Pt/Ru/Pt insertion multilayer is performed, which creates an in-plane asymmetry to generate interlayer DMI. The existence of interlayer DMI is demonstrated by the asymmetric magnetization switching behavior in an angular field sweeping measurement. Macro-spin simulations reveal the SAF dynamics under different SOT scenarios. With the existence of chiral spin textures between top and bottom magnetic layers, the SOT induced in-plane alignment of magnetizations favors a certain out-of-plane splay, which result in deterministic switching after spin precession. Guided by the theoretical prediction, field-free SOT switching is experimentally achieved, where we observe the consistency of effective switching field and the interlayer DMI strength observed in anomalous Hall measurements.

## Results

### Sample preparation and characterization of interlayer DMI

After a careful optimization process, the film studied here has a structure of Si/SiO2//Ta(2nm) /Pt(5nm) /Co$_{80}$B$_{20}$(0.6nm) /Pt(30°,0.6nm) /Ru(30°,0.8nm) /Pt(30°,0.6nm) /Co$_{80}$B$_{20}$(0.6nm) /Pt(2nm) deposited on 100 nm thermally oxidized Si by direct current (DC) magnetron sputtering (details in Method). The angles between brackets indicate the angle between deposition direction and the substrate normal, where 0° is used when not specified. The in-plane azimuthal direction of asymmetric sputtering is defined as $\varphi = 0°$ in this paper. See Fig. S1 for a schematic of tilted sputtering. The asymmetric Pt insertion layers provide strong interlayer DMI (31) and the Ru layer provides the Ruderman-Kittel-Kasuya-Yosida (RKKY) interaction that antiferromagnetically couples the two FM layers (34). The samples are then patterned into Hall bars along different azimuthal directions using E-Beam Lithography and a standard lift-off process. The 2nm Pt capping layer is considered to be partially oxidized and the bottom 5nm Pt provides the majority of SOT in the switching experiments.

The interlayer DMI energy between $\boldsymbol{m_1}$ and $\boldsymbol{m_2}$ can be written as (35)

$$\mathrm{E}_{DMI} = -\boldsymbol{D_{1,2}} \cdot (\boldsymbol{m_1} \times \boldsymbol{m_2}) \tag{1}$$

where $\boldsymbol{m_1}$ and $\boldsymbol{m_2}$ are unit vectors of magnetization in top and bottom magnetic layers (FM1 and FM2) and $\boldsymbol{D_{1,2}}$ is a DMI vector along ±x, depending on the orientation of the structural asymmetry on the y axis in the geometry of Fig.1.

To verify the aforementioned interlayer DMI in our SAF sample, we measure the asymmetric switching behavior through anomalous Hall effect (AHE) measurement (29). Figure 1B and Fig. 1C show the example of the switching of $\boldsymbol{m_2}$ from down to up while $\boldsymbol{m_1}$ remains up. When $\boldsymbol{m_2}$ is switched by an external field $\boldsymbol{H_{ext}}$, the perpendicular component $H_z$ drives the switching and the in-plane component $H_\parallel$ helps to define the in-plane switching direction. The DMI energy in Eq. (1) is equivalent to a field $H_{iDMI}$ acting on $\boldsymbol{m_2}$ along the direction of $\boldsymbol{D_{1,2}} \times \boldsymbol{m_1}$ and, depending on the sign of $\boldsymbol{D_{1,2}}$ and the direction of $H_\parallel$, assisting or hindering the upward switch of $\boldsymbol{m_2}$. Therefore, the sign and magnitude of the DMI can be extracted from the asymmetric switching behavior driven by $\boldsymbol{H_{ext}}$ with $H_\parallel$ along different directions. As schematically represented in Fig. 1D, we measure the magnetization switching by sweeping the elevation angle (θ) of $H_{ext}$ for different values of the azimuthal angle $\varphi$, where $H_z = H_{ext} \cos\theta$ and $H_\parallel(\varphi) = H_{ext} \sin\theta$. We use a fixed value of external field $H_{ext} = 80$ mT, which is slightly larger than the out-of-plane saturation field of the sample. The anomalous Hall resistance ($R_{AHE}$) curves with $\varphi = 0°$ and $\varphi = 180°$ are shown in Fig. 1E. A loop shift $\Delta|H_{SW}|$ is observed, which demonstrates the asymmetric switching behavior in this structural asymmetric direction. We summarized the switching field $H_{SW}$ of both layers, defined by $H_z$ which causes the $R_{AHE}$ to change 50% during the entire switching, as shown in Fig 1F and Fig. 1G. We can see that $\Delta|H_{SW}| = H_{SW}(\varphi) - H_{SW}(\varphi + 180°)$ reaches its maximum value at $\varphi = 0°$, and becomes negligible at $\varphi = 90°$, consistent with former reports (29, 35). This confirms that the interlayer DMI is indeed a result of the structural asymmetry along $\varphi = 0°$ induced by the tilted sputtering.

**Mechanism of SAF field-free SOT switching**

Next, we show how the interlayer DMI enables field-free SOT switching of SAFs. The evolution of a SAF spin texture during the application of a SOT current pulse is illustrated in a macro-spin model in Fig. 2A to Fig. 2D. The initial spin texture before a SOT current pulse is shown in Fig. 2A. As we demonstrate later in this paper, when we apply a large SOT current along -x to a SAF without interlayer DMI, $\boldsymbol{m_1}$ and $\boldsymbol{m_2}$ will be aligned by SOT toward the in-plane spin polarization along -y, as shown in Fig. 2B. Under this configuration, the introduction of an interlayer DMI vector $\boldsymbol{D_{1,2}}$ along +x will slightly splay $\boldsymbol{m_1}$ and $\boldsymbol{m_2}$ and give them a perpendicular component, as shown in Fig. 2C. After the SOT pulse is over, this initial perpendicular component will lead to deterministic switching through spin precession, as shown in Fig. 2D. The SAF could thus be switched by a SOT current pulse under zero external field.

Fig. 2E illustrates the field-free SOT switching under a domain nucleation and propagation regime as expected in our 20 μm scale Hall bar device. With the help of the chiral spin texture, a large in-plane spin current gives $\boldsymbol{m_1}$ and $\boldsymbol{m_2}$ out-of-plane components in opposite directions, which initiates domain nucleation in the SAF. With the existence of SOT, the domain walls (DWs) of both layers favor Bloch type. This results in the magnetization of top and bottom DWs to be splayed and DWs to be driven to expand by the interlayer DMI under a similar regime. Therefore, in a μm scale Hall bar device, the SAF will ultimately be switched following a domain nucleation and propagation process. The deterministic switching direction depends on the direction of spin polarization and interlayer DMI.

**Macro-spin simulations.**

To demonstrate the proposed mechanism, we need to first answer a question: How could SOT efficiently align the magnetizations in SAF toward the in-plane spin polarization? Considering the damping-like SOT, the spin dynamics follow the Landau-Lifshitz-Gilbert (LLG) equation:

$$\frac{\partial \boldsymbol{m}}{\partial t} = -\gamma \mu_0 \left( \boldsymbol{m} \times \boldsymbol{H}_{eff} \right) + \alpha \left( \boldsymbol{m} \times \frac{\partial \boldsymbol{m}}{\partial t} \right) + \gamma \mu_0 H_{SOT} \boldsymbol{m} \times (\boldsymbol{\sigma} \times \boldsymbol{m}) \qquad (2)$$

where $\gamma$, $\mu_0$ and $\alpha$ are the gyromagnetic ratio, magnetic permeability in vacuum and damping factor, respectively. $\boldsymbol{m}$ is a normalized vector for magnetization and $\boldsymbol{\sigma}$ is a normalized vector for spin polarization. $\boldsymbol{H}_{eff}$ is an effective field including magnetic anisotropy, exchange coupling and interlayer DMI. We note that although the introduction of a field-like torque will influence the magnetization dynamics and the switching speed (36–38), it will not affect the main conclusion of the simulation in this paper. See details in supplementary material.

We first consider the SOT induced dynamics in a single FM layer. In this case, $\boldsymbol{H}_{eff} = H_k \cos\theta \, \hat{\boldsymbol{z}}$, where $H_k$ is the anisotropy field and $\hat{\boldsymbol{z}}$ is a unit vector along the z direction. We investigate the spin dynamics induced by a SOT current along +x, which results in a spin polarization $\boldsymbol{\sigma}$ along +y under a positive spin Hall angle. When the amplitude of SOT is relatively small, the SOT term in Eq. (2) $\gamma \mu_0 H_{SOT} \boldsymbol{m} \times (\boldsymbol{\sigma} \times \boldsymbol{m})$ can be approximated by $\gamma \mu_0 \boldsymbol{m} \times H_{SOT} \hat{\boldsymbol{x}}$. Due to an interplay of SOT and anisotropy field, the magnetization will precess towards a position tilted in the x, as shown in Fig. 3A. However, when the SOT current density $J_{SOT}$ is above a critical value $J_c$, this spin precession is no longer stable. Instead, it will directly lead $\boldsymbol{m}$ toward a new equilibrium orientation along +y, where both SOT and anisotropy torques are 0. The spin precession of this process is shown in Fig. 3B. We summarize the evolution of the final equilibrium orientation with varying $J_{SOT}$ in Fig. 3C, where $J_{SOT}$ is swept from $0 \text{ MA/cm}^2$ to $100 \text{ MA/cm}^2$ with a step of $3.33 \text{ MA/cm}^2$. We can see that under zero external field, a $J_{SOT}$ larger than $J_c$ can efficiently align $\boldsymbol{m}$ into the in-plane direction toward the spin polarization.

With an exchange coupling field in the SAF, SOT induced magnetization behavior is more complex. It has been demonstrated that the SAF could be switched by SOT in different symmetry breaking scenarios (i) Spin injection with mirror symmetry relative to the x-y plane ($C_{xy}$) (13, 35). (ii) Spin injection with broken $C_{xy}$ symmetry (14–17). We first explore the SAF dynamics of both SOT scenarios under zero external field and with zero interlayer DMI. The effective field of the antiferromagnetic RKKY coupling for $\boldsymbol{m_{1(2)}}$ could be written as $\mu_0 \cdot \boldsymbol{H_{ex1(2)}} = -\frac{J_{RKKY}}{M_S \cdot t_{FM}} \hat{\boldsymbol{m}}_{2(1)}$, where $J_{RKKY}$, $M_S$, $t_{FM}$, are the energy density of antiferromagnetic RKKY coupling, saturation magnetization and FM thickness, respectively (See Table S1 for details). $\hat{\boldsymbol{m}}_{1(2)}$ is a unit vector for the top (bottom) magnetization. As compared in Fig. 3D and Fig 3F, spin injection with broken $C_{xy}$ symmetry could efficiently align $\boldsymbol{m_1}$ and $\boldsymbol{m_2}$ toward $\boldsymbol{\sigma}$, while spin injection with $C_{xy}$ symmetry stabilizes the magnetization precession and prevents the in-plane alignment of magnetizations. We also observe in Fig. 3E that the spin injection only in the bottom layer, which is expected to be the case in our sample, could also align $\boldsymbol{m_1}$ and $\boldsymbol{m_2}$ toward $\boldsymbol{\sigma}$. The duration of spin precession is longer compared with Fig. 3D. We note that since $\boldsymbol{m_1}$ is purely driven by the RKKY exchange field in this SOT scenario, the RKKY strength should be larger than $H_k$ to overcome the anisotropy. However, if the RKKY coupling is too strong, it will prevent the the two magnetization vectors to co-align along +y, and the system will end up in a stable oscillation with an opening angle between $\boldsymbol{m_1}$ and $\boldsymbol{m_2}$. See supplementary material for details.

The simulation also shed light on understanding the conventional in-plane field assisted SAF SOT switching. As shown in Fig. 3F, when the top and bottom layers have opposite signs of spin Hall angle, the exchange coupling will stabilize the spin precession and prevent the magnetizations overcoming their energy barriers. This explains the large external field required for SOT switching of SAF in existing literature (13, 35). On the other hand, during the field assisted SOT switching of SAF with broken $C_{xy}$ symmetry, the top and bottom magnetizations are aligned to a same in-plane direction under zero external field so they will favor the same out-of-plane direction under a certain external field. Therefore, the magnetic switching will rely on the small remanent magnetization to determine its direction. This decreases the robustness of magnetization operation. The shortcomings in both scenarios have greatly hindered the development of field-assisted SAF SOT devices, which are solved in our strategy.

Next, we examine the function of an interlayer DMI in SAF SOT dynamics. Based on Eq. (1), the interlayer DMI effective field for $\boldsymbol{m_{1(2)}}$ can be written as:

$$\mu_0 \cdot \boldsymbol{H_{iDMI1(2)}} = \pm \frac{1}{M_S \cdot t_F} \cdot \hat{\boldsymbol{m}}_{2(1)} \times \boldsymbol{D_{1,2}} \qquad (3)$$

The $m_z$ evolution of top and bottom layers without an interlayer DMI in Fig. 3E is summarized in Fig. 3G, where $m_z$ is the z component of magnetization. When we introduce an interlayer DMI vector $\boldsymbol{D_{1,2}} \parallel \boldsymbol{J_{SOT}}$ along +x, the in-plane alignment of magnetizations will give opposite perpendicular components to both layers. After the SOT pulse, deterministic switching can be achieved through spin precession, as shown in Fig. 3H. When $\boldsymbol{D_{1,2}} \perp \boldsymbol{J_{SOT}}$, however, the magnetizations do not have out-of-plane components and will precess to a random direction after the SOT pulse, as shown in Fig. 3I. The simulation result is consistent with the proposed field-free switching mechanism.

**SOT switching experiment under zero external field.**

Finally, we present the results of the field-free switching experiment. Figure 4A shows the normalized polar magneto-optical Kerr (p-MOKE) signal of our SAF sample while varying $H_z$. Due to the larger Kerr signal from the top FM layer, we can distinguish the two different anti-parallel magnetization states. We then apply a 100 ns current pulse along -x, which means $\boldsymbol{J_{SOT}} \parallel \boldsymbol{D_{1,2}}$ in our Hall bar device. The domain structure after pulse application as observed by Kerr microscopy is shown in Fig 4b. 80% of the Hall bar area is switched under zero external field, where the unswitched part on the edge is generally considered to be hindered by the Oersted field generated by the charge current (39). The $\boldsymbol{D_{1,2}}$ inferred from the switching behavior is consistent with the previous AHE measurements. We summarize in Fig. 4C the switching curves of different Hall bar devices with varying orientation of the channel track respect to the orientation of $\boldsymbol{D_{1,2}}$. Each data point is an average of 3~7 Kerr images. (See Fig. S4 for the representative Kerr images of different device angles). We can see that the switching area reaches its maximum when $\boldsymbol{D_{1,2}} \parallel \boldsymbol{J_{SOT}}$ and negligible switching is observed when $\boldsymbol{D_{1,2}} \perp \boldsymbol{J_{SOT}}$, consistent with the theoretical prediction.

To further verify the origin of the observed field-free switching behavior in our experiment, we compare the $H_{iDMI}$ derived from the SOT switching and AHE measurement. As shown in the inset of Fig. 5 the $H_{iDMI}$ that drives the switching of $\boldsymbol{m_1}$ is partially canceled by the Oersted field. Therefore, $H_{iDMI}$ is comparable with the Oersted field at the boundary of the switched area. We calculate the Oersted field on the average switching boundary generated by a 90 MA/cm$^2$ current pulse, as shown by the blue points and fitting in Fig. 5. This represents the $H_{iDMI}$ on FM1 during the switching experiment. See supplementary material for the calculation of the Oersted field. On the other hand, in the AHE measurement, $H_{iDMI}$ could be characterized by the loop shift $\Delta|H_{SW}|$. $\Delta|H_{SW}|$ of FM1 is summarized from Fig. 1G and depicted as the red points and fitting in Fig. 5. From the comparison of the two curves, the interlayer DMI strength derived from switching and AHE measurements show a similar trend, where by using a sine fitting, amplitudes of 1.99 mT and 0.89 mT, respectively are found. The unequal $H_{iDMI}$ values can be explained by the different magnetization configurations in the two experiments. The equal trends in Fig. 5 strongly support that the observed field-free switching is indeed derived from interlayer DMI.

**Discussion**

The observations in this paper may open up a new route towards SAF-based high performance SOT devices. Compared with conventional field-free switching solutions, the proposed strategy engineers only the middle insertion layer to create the asymmetry. Therefore, it is compatible with most of the well demonstrated high performance device structures, such as CoFeB/MgO/CoFeB based magnetic tunnel junctions (MTJs) with high TMR (6), β-W (18, 36), two dimensional material (40) and topological insulators (41) with high current-spin conversion efficiency. Moreover, an interlayer DMI does not require uneven distribution of layer thickness, which preserve the large-scale wafer uniformity required by mass production. (See details in supplementary material) We note that further studies to create a stronger interlayer DMI alternatively, by magnetic asymmetry of the insertion layer (42), for example, will benefit the mass production and increase the switching robustness in the industrial applications.

In summary, we theoretically propose and experimentally achieve the field-free SOT switching of a SAF through an interlayer DMI parallel to the electric current. Interlayer DMI is demonstrated by azimuthal angular dependent AHE measurement in our sample. By macro-spin simulation, SOT induced SAF dynamics with different symmetry scenarios are studied. We find that SOT with broken $C_{xy}$ symmetry can efficiently align the magnetizations towards the spin polarization direction, which could lead to deterministic switching initialized by an interlayer DMI. The simulation also contributes to the understanding of a high external field required in traditional field-assisted SAF SOT switching. SOT switching experiments demonstrate the proposed mechanism and show consistent $H_{iDMI}$ with the AHE measurement, which proves that the field-free switching is indeed derived from the interlayer DMI. This systematic study of SAF field-free SOT switching may pave a way for the next generation ultrafast, high density and energy efficient spintronic devices.

**Materials and Methods**

**Sample preparation**

Sputtered magnetic multilayers Ta(2nm) /Pt(5nm) /Co$_{80}$B$_{20}$(0.6nm) /Pt(30°,0.6nm) /Ru(30°,0.8nm) /Pt(30°,0.6nm) /Co$_{80}$B$_{20}$(0.6nm) /Pt(2nm) were deposited on 100 nm oxidized Silicon substrates under room temperature using 20 W direct-current (DC) magnetron sputtering. The base pressure and working pressure are 1×10$^{-6}$ Pa and 1 Pa, respectively. No post-annealing is performed. To fabricate the Hall bar devices, an E-Beam Lithography and lift-off process is used. The width and length of the current channel of Hall bar are 20 μm and 140 μm, respectively.

**Macro-spin simulation**

The numerical macro-spin simulation is done in Matlab at zero degree Kelvin. The main parameters are summarized in table S1. The SAF dynamics shown in Fig. 3d to Fig. 3g are simulated with the SOT turned on for 1.5 ns, whereas the SAF dynamics in Fig. 3h and Fig. 3i are simulated for 4 ns with the SOT turned on in the first 1.5 ns. As shown in Fig. 3i, after the SOT current is switched off, $m_1$ and $m_2$ precess back to their initial directions. We note that this random direction of switching could change when the pulse duration changes slightly in our simulation. The effective field of RKKY coupling is derived from the following expression of RKKY energy per unit area:

$$E_{RKKY} = J_{RKKY} \, m_1 \cdot m_2$$

where a positive $J_{RKKY}$ corresponds to antiferromagnetic RKKY coupling.

**Experimental setup**

For the anomalous Hall resistance measurements, we use Keithley model 6221 current source to apply a 1 mA, 1 kHz alternating current and a Stanford Research SR830 lock-in amplifier to measure the Hall voltage. A 5mm*5mm film sample is studied, with the wire bonding for electric connection performed near the 4 corners of the square sample. The SOT driven magnetization switching is studied by an Evico Kerr microscope in polar mode, which shows a high-resolution digital imaging of the out-of-plane magnetization component of the top magnetic layer in SAF. The magnetization direction could be judged from the brightness of the digital point relative to a critical value. The percentage of region where the magnetization of the top layer points up is counted after we blur the Kerr image with 9*9 nearby points. Besides, AVTECH model AVI-V-HV2A-B pulse generator is used for the application of 100ns current pulses.

**Acknowledgments**

This work was supported in part by National Natural Science Foundation of China (Grant No. 92164206), International Collaboration Project (Grant No. B16001), Beihang Hefei Innovation Research Institute Project (Grant No. BHKX-19-02), China Scholarship Council (CSC) and the European Union's Horizon 2020 research and innovation program under the Marie Skłodowska-Curie grant agreement (860060). Zilu Wang gratefully thank Daoqian Zhu, Mark De Jong,

**Figures and Tables**

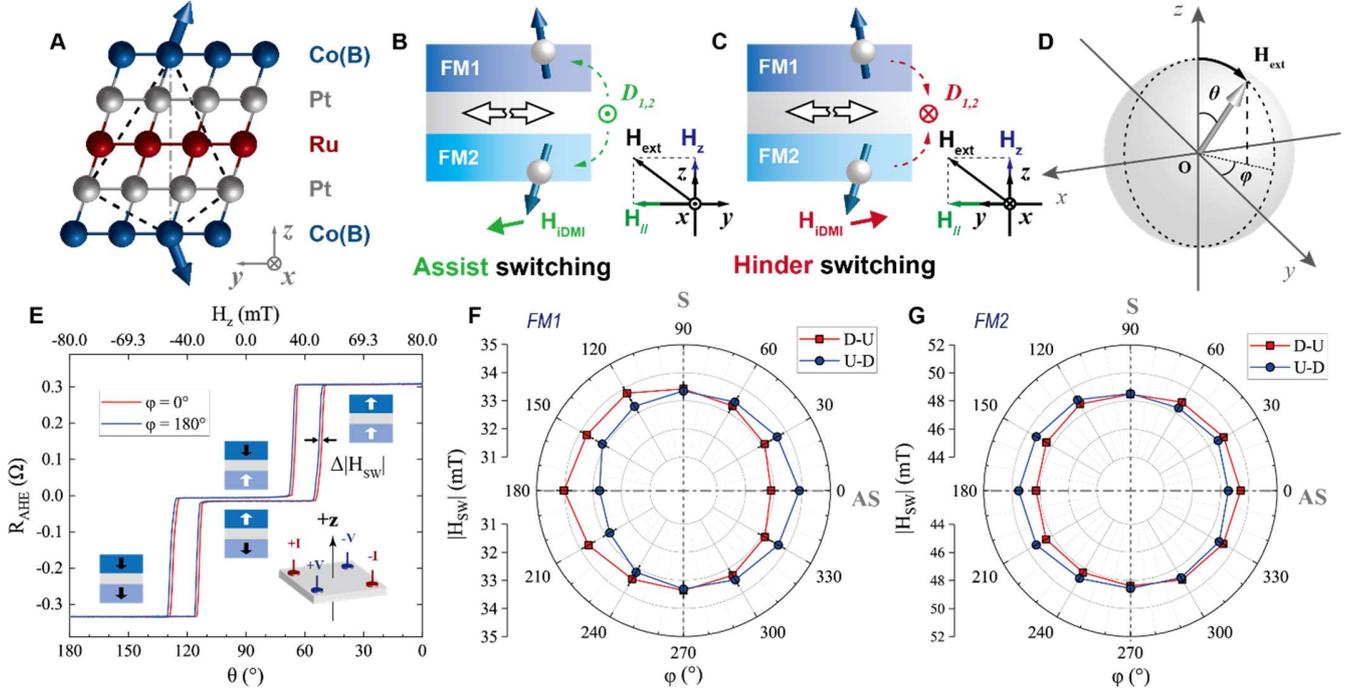

**Figure 1. Asymmetric azimuthal dependent switching behavior induced by interlayer DMI.**
**(A)** Schematic of chiral spin texture caused by interlayer DMI, illustrated on a simplified atomic structure of our sample. **(B-C)** Field induced switching of magnetization assisted (B) or hindered (C) by interlayer DMI. White arrows show the direction of the structural asymmetry. The circle with a cross (point) indicates the DMI vector, into (out from) the plane of the drawing. **(D)** Elevation angular sweep of external field in the azimuthal dependent switching experiment, illustrated in a polar coordinate system. **(E)** Comparison of anomalous Hall resistance under an 80mT rotating external field along $\varphi = 0°$ and $\varphi = 180°$. The top axis shows the z component of the field. The loop shift $\Delta|H_{SW}|$ indicates the asymmetric switching behavior induced by the interlayer DMI. Inset show the contact configuration of the Hall measurements, with the elevation angle of external field $\theta_B = 180°$. **(F-G)** Summary of up to down (U-D) and down to up (D-U) switching field $|H_{SW}|$ of FM1 (F) and FM2 (G) as a function of $\varphi$.

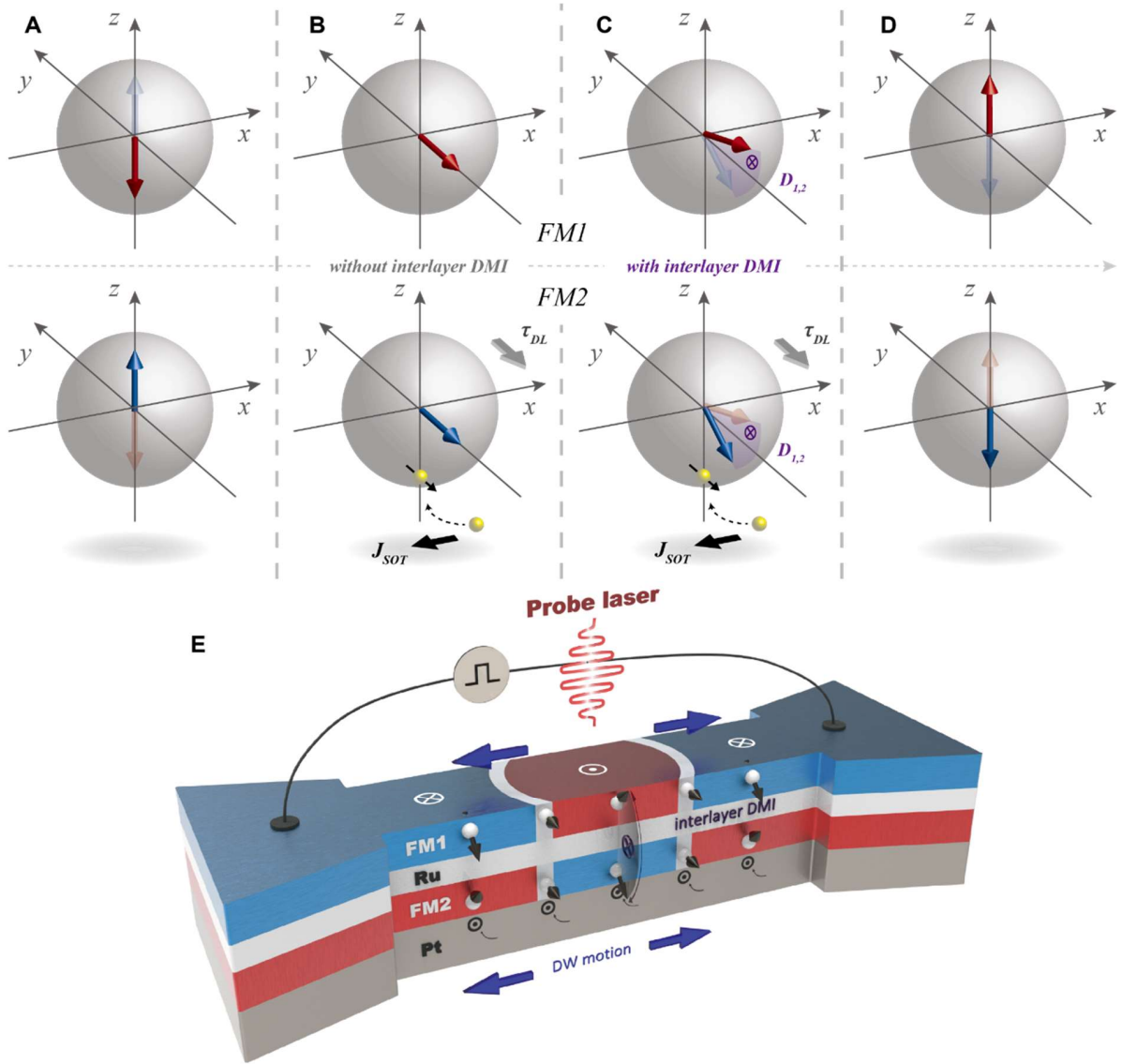

**Figure 2. Schematic illustration of SAF field-free SOT switching through interlayer DMI. (A-D)** Spin textures' evolution with time during the application of a SOT current pulse along -x. Magnetizations of the top and bottom magnetic layers in SAF are shown in the top and bottom spheres, respectively. (A) Initial magnetization state before SOT pulse. (B-C) Magnetization states during application of SOT in SAF with no interlayer DMI (B) and an interlayer DMI $D_{1,2}$ along +x (C). $\tau_{DL}$ indicates the damping-like term of SOT. (D) Final state after application of SOT. **(E)** A sketch of the experimental setup where the switching mechanism under a domain nucleation and propagation regime is indicated. The probe laser is used to detect the MOKE signal in our experiment. The purple cross surrounded by a circle represents the interlayer DMI $\boldsymbol{D_{1,2}}$ between two layers.

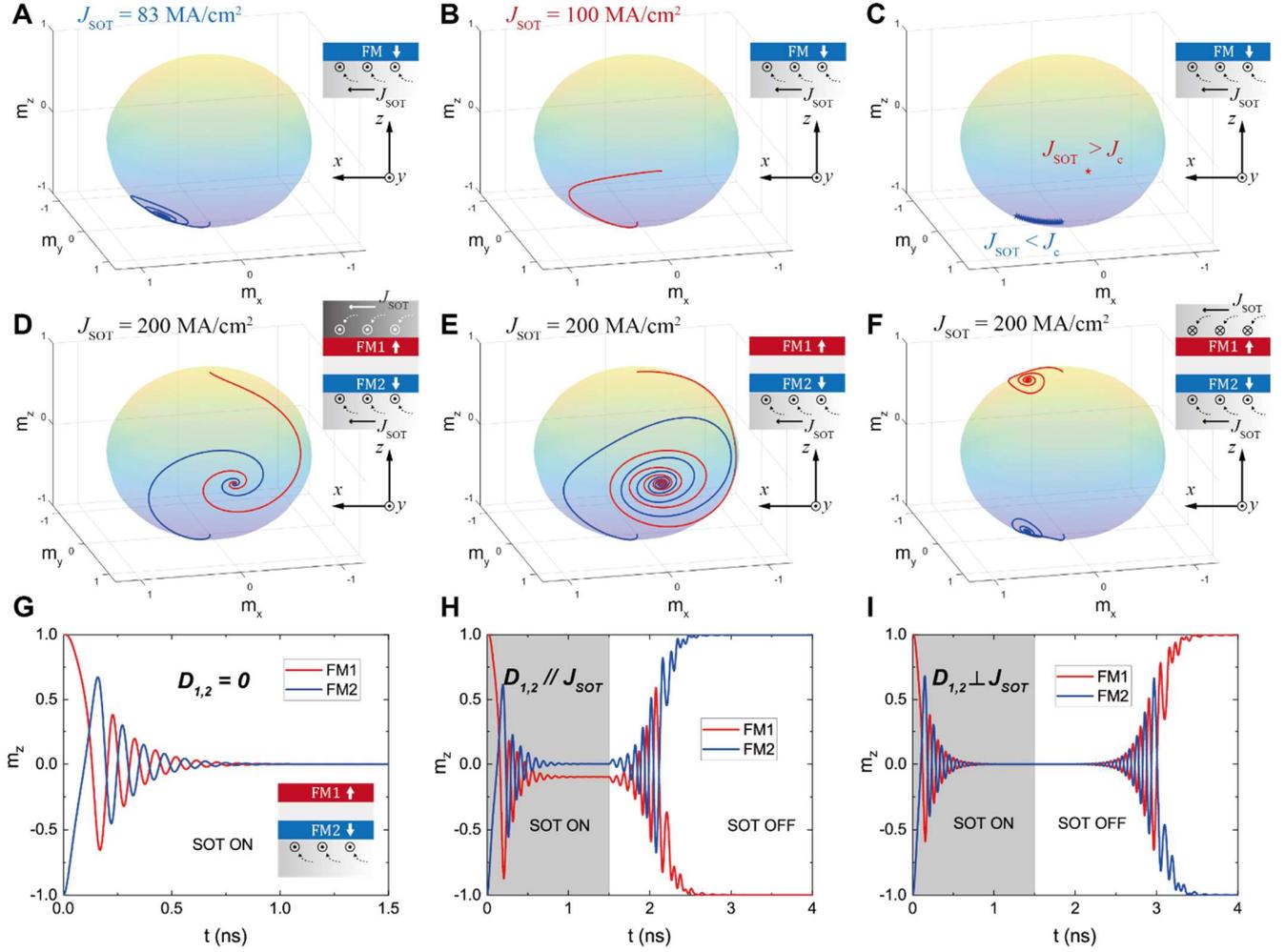

**Figure 3. Macro-spin simulation of SOT induced magnetization dynamics. (A-B)** Spin dynamics in FM single layer induced by SOT below (A) and above (B) a critical value $J_c$. **(C)** Summary of ending positions of precession as a function of $J_{SOT}$. **(D-F)** Spin dynamics of SAF induced by SOT from the top and bottom layers with opposite spin Hall angle (D), SOT from only bottom layer with spin Hall angle (E), SOT from the top and bottom layers with same spin Hall angle (F). **(G-I)** z component magnetization $m_z$ dynamics with no interlayer DMI (G), interlayer DMI parallel to $J_{SOT}$ (H) and interlayer DMI perpendicular to $J_{SOT}$ (I). See Table S1 for detailed parameters used in the simulations.

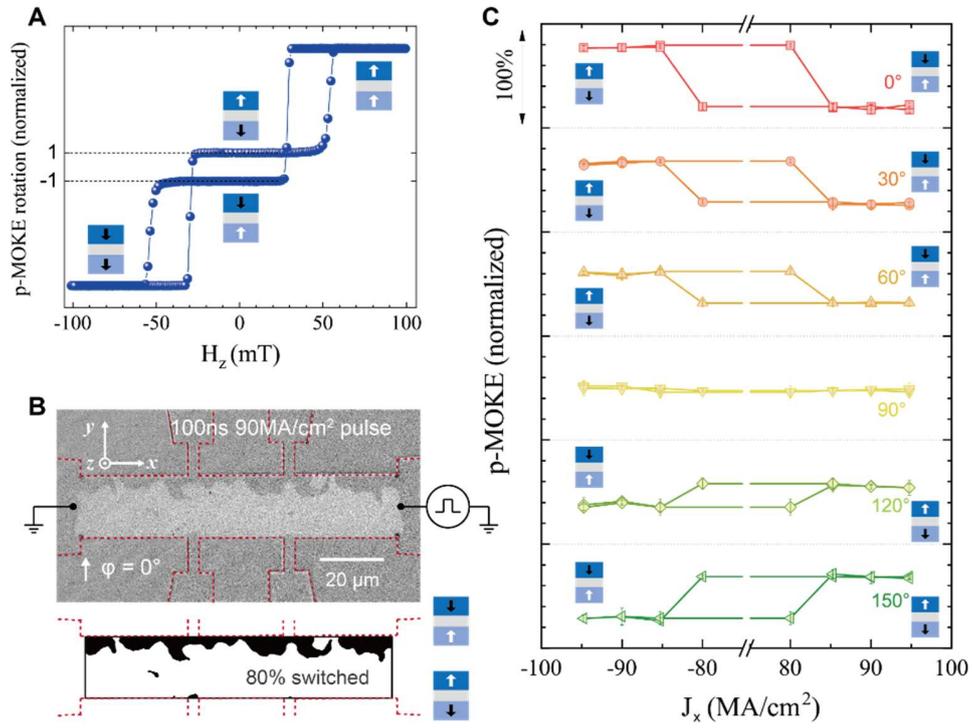

**Figure 4. Field-free SOT switching detected by optical Kerr measurement. (A)** Normalized p-MOKE signal of the studied film. **(B)** Kerr image after the application of a current pulse. The switching area is counted from the composite image below. **(C)** Summarization of current induced switching behavior as a function of φ. As shown in (B), the direction of φ is along the voltage channel of the Hall bar device.

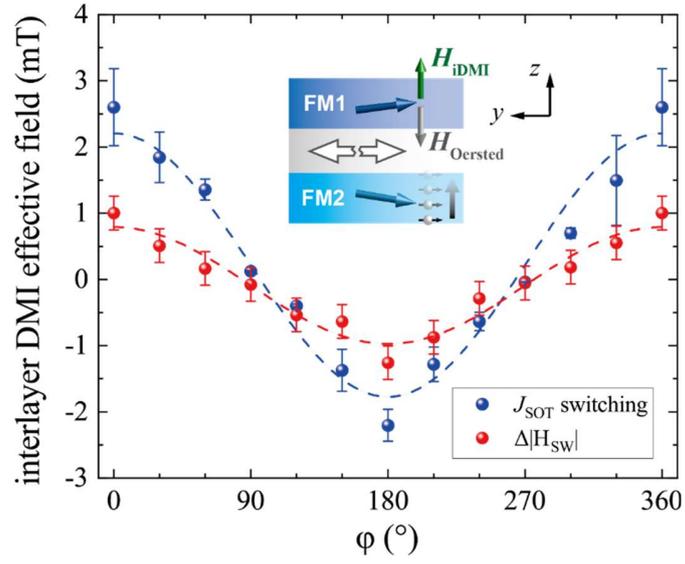

**Figure 5. Effective field of interlayer DMI, derived from AHE and SOT switching measurement, respectively.** The blue points show the Oersted field at the average boarder between the switching and non-switching regions under 90 MA/cm² current pulse. The red points show $\Delta|H_{SW}|$ of FM1 summarized from AHE experiment. The dotted lines are the fittings using a sine function.

**Supplementary Information for**

Field-free spin orbit torque switching of synthetic antiferromagnet through interlayer Dzyaloshinskii-Moriya interaction

Zilu Wang[a,b,*], Pingzhi Li[b], Yuxuan Yao[a], Youri L.W. Van Hees[b], Casper F. Schippers[b], Reinoud Lavrijsen[b], Albert Fert[a,c], Weisheng Zhao[a,*], and Bert Koopmans[b]

*Zilu Wang, Weisheng Zhao.

Email: z.wang1@tue.nl; weisheng.zhao@buaa.edu.cn

**This PDF file includes:**

Supplementary text

Fig. S1 Thickness distribution during the slanted epitaxial film growth.

Fig. S2 Macro-spin simulation of SOT induced magnetization dynamics with different RKKY coupling energy.

Fig. S3 Macro-spin simulation of SOT induced magnetization dynamics with different field like torque ratios.

Fig. S4 Representative Kerr images of SOT switching in different azimuthal directions.

Fig. S5 Schematic illustration of Oersted field generated by evenly distributed current in Hall bar.

Table S1. Main parameters for simulation

**Thickness distribution during the slanted epitaxial film growth.**

We will demonstrate here that the structural asymmetry brought by the tilted sputtering is mainly contributed by the internal crystal asymmetry. The uneven distribution of the in-plane thickness is comparably negligible. Fig. S1a and Fig. S1b illustrate the difference between the face to face sputtering and the tilted sputtering. As shown in Fig. S1b, an atomic structural asymmetry is generated during the tilted sputtering. However, the distance between the sample and target is not even throughout the wafer, which brings an uneven distribution of layer thickness. We calibrate the thickness distribution of tilted sputtered Ru. A large tilt angle of 50° is adopted to make the thickness distribution clearer for observation. Layer thicknesses at different positions are characterized by atomic force microscope (AFM). As summarized in Fig. 1c, less than 12% thickness difference is observed for 1cm displacement along the asymmetric direction. Therefore, the thickness distribution is smaller than 0.12% per 100µm in the studied sample. We conclude that the thickness distribution is a trivial factor in generating interlayer DMI, which guarantee the large-scale wafer uniformity of this field-free solution.

**Influence of RKKY strength on SOT driven SAF dynamics**

When we apply a SOT with spin injection only in $m_2$, the dynamics of $m_1$ is purely driven by the RKKY coupling. We simulate in Fig. S2 the SAF dynamics with different RKKY strength without interlayer DMI. As shown in Fig. S2a and Fig. S2b, when the RKKY coupling energy $J_{RKKY}$ = -0.05×$10^{-3}$ J/$m^2$, it is not strong enough to overcome the anisotropy field. In this case, the spin injection in $m_2$ only exerts a small perturbation to $m_1$ and is not able to align $m_1$ toward the spin polarization. When the RKKY coupling energy increases to $J_{RKKY}$ = -0.2×$10^{-3}$ J/$m^2$, $m_1$ and $m_2$ will precess with a decaying amplitude, and end up with alignment toward the spin polarization along +y, as shown in Fig. S2c and Fig. S2d. As the RKKY coupling energy further increase to $J_{RKKY}$ = -0.5×$10^{-3}$ J/$m^2$, however, the antiferromagnetic coupling in the SAF is too strong to be overcome by the SOT. Therefore, after first precessing with a decaying amplitude, the spin precession ends up in continuous precession around the spin polarization direction +y, as shown in Fig. S2e and Fig.S2f.

Therefore, we conclude that a proper RKKY coupling energy is required to enable the proposed field-free switching mechanism driven by single layer spin injection.

**Influence of field-like torque on SOT driven SAF dynamics**

Considering the field-like torque, the spin dynamics follow the Landau-Lifshitz-Gilbert (LLG) equation:

$$\frac{\partial \boldsymbol{m}}{\partial t} = -\gamma \mu_0 \left( \boldsymbol{m} \times \boldsymbol{H_{eff}} \right) + \alpha \left( \boldsymbol{m} \times \frac{\partial \boldsymbol{m}}{\partial t} \right) + \gamma \mu_0 H_{SOT}^{DL} (\boldsymbol{m} \times \boldsymbol{\sigma} \times \boldsymbol{m}) + \gamma \mu_0 H_{SOT}^{FL} (\boldsymbol{m} \times \boldsymbol{\sigma}) \text{ (S1)}$$

where $H_{SOT}^{DL}$ and $H_{SOT}^{FL}$ are the effective fields for the damping-like and field-like term of SOT, respectively. We define the field like torque ratio $\eta = H_{SOT}^{FL}/H_{SOT}^{DL}$. As shown in Fig. S3a to Fig. S3c, the introduction of a field-like torque causes asymmetric oscillation between top and bottom layers, whereas the effective damping factor depends on the sign relation between field-like and damping-like torques. As shown in Fig. S3d to Fig. S3f the field-free switching assisted by interlayer DMI could be achieved for both $\eta = -1$, $\eta = 0$ and $\eta = +1$. However, the field-like torques can influence the robustness and speed of the switching process.

**Calculation of the Oersted field at the Hall bar edge.**

To estimate $H_{iDMI}$ during SOT switching, we need to calculate the current induced Oersted field at the boundary of the switching area. As shown in Fig. S5, the Oersted field generated by the grey region is symmetric relative to the calculated point. Therefore, it does not contribute to the net perpendicular Oersted field. The Oersted field at the calculation point generated by the yellow zone can be calculated through the Ampère's law:

$$H_{Oe} = \int_{(1-p)\cdot w}^{p\cdot w} \frac{\mu_0 J t_{FM}}{2\pi x} dx$$

where $J$, $p$, and $w$ are the current density, switching percentage and the Hall bar width, respectively.

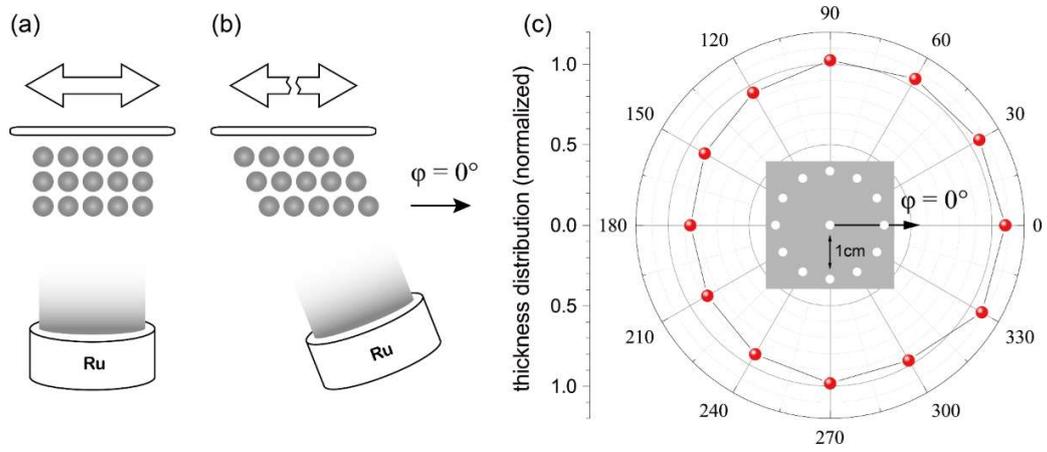

**Fig. S1 Thickness distribution during the slanted epitaxial film growth. a-b** Schematic illustration of face to face (a) and slanted (b) magneton sputtering. **c** Thickness distribution among the sample, normalized by the thickness of the center point. Inset shows the distribution of test points.

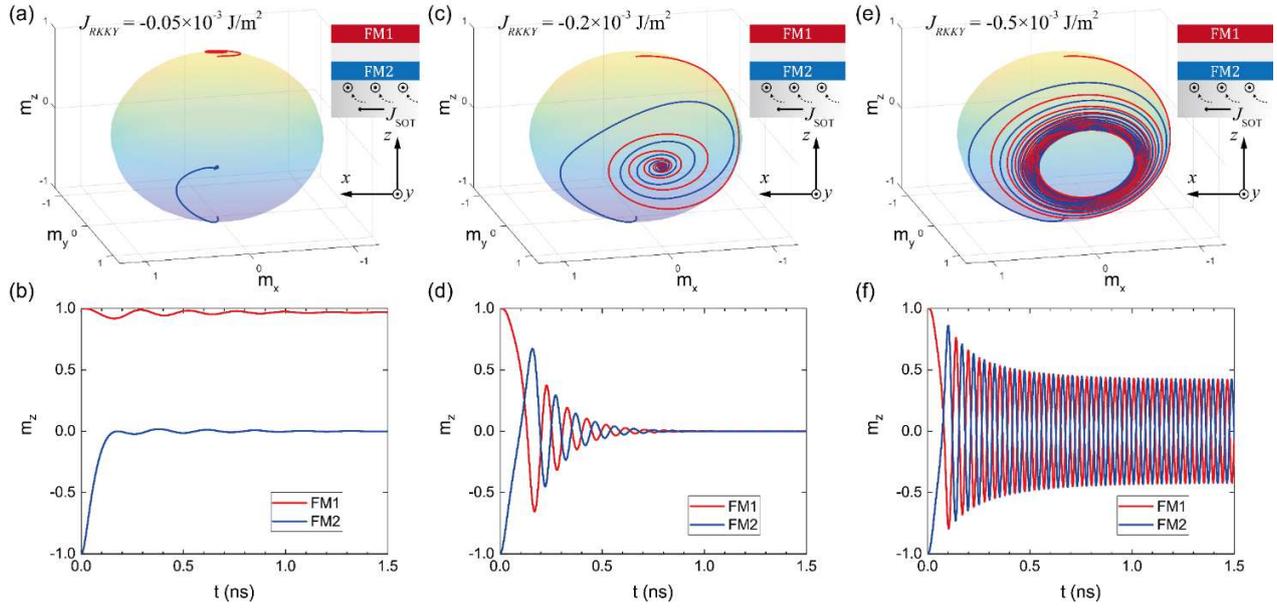

**Fig. S2 Macro-spin simulation of SOT induced magnetization dynamics with different RKKY coupling energy.** 200 MA/cm² SOT current is applied and the spin injection is in only the bottom layer. **a-b** SAF precession with $J_{RKKY}$ = -0.05×10⁻³ J/m². **c-d** SAF precession with $J_{RKKY}$ = -0.2×10⁻³ J/m². **e-f** SAF precession with $J_{RKKY}$ = -0.5×10⁻³ J/m².

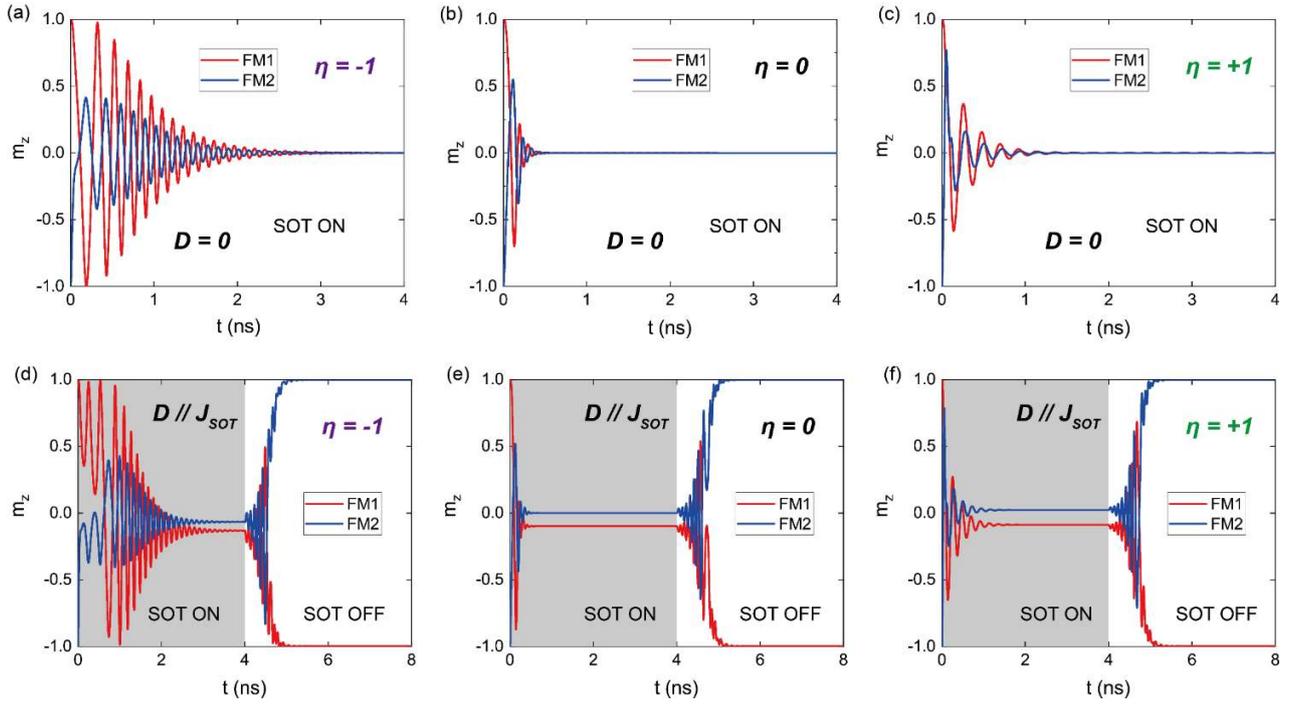

**Fig. S3 Macro-spin simulation of SOT induced magnetization dynamics with different field like torque ratios. a-c** Spin dynamics of SAF without interlayer DMI induced by spin injection in bottom layer generated from a 300 MA/cm² DC current. The field like torque ratios are $\eta = -1$ **(a)**, $\eta = 0$ **(b)** and $\eta = +1$ **(c)**. **d-f** Field-free SOT switching of SAF induced by spin injection in bottom layer generated from a 4ns 300 MA/cm² current pulse, with the help from an interlayer DMI $\boldsymbol{D} \parallel \boldsymbol{J_{SOT}}$. The field like torque ratios are $\eta = -1$ **(d)**, $\eta = 0$ **(e)** and $\eta = +1$ **(f)**.

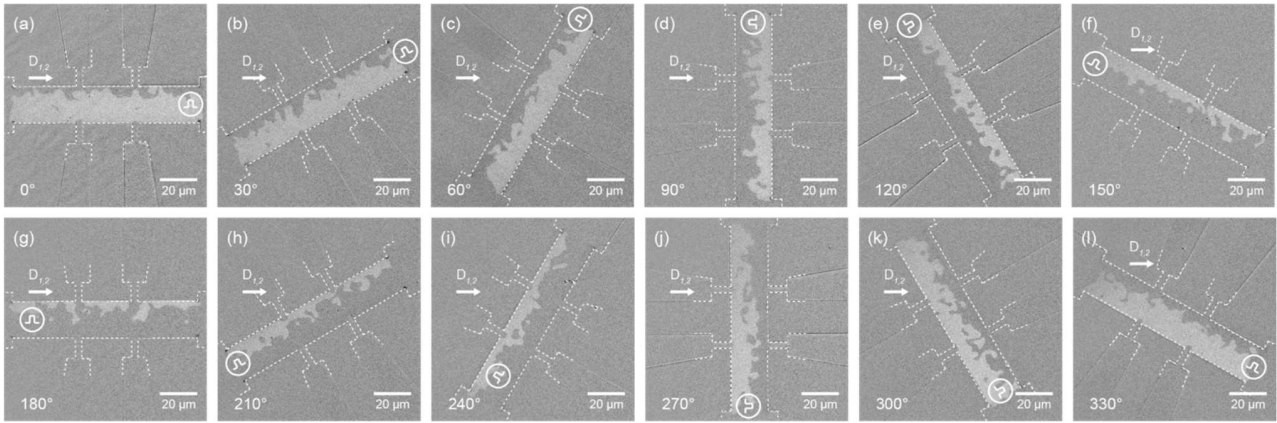

**Fig. S4 Representative Kerr images of SOT switching in different azimuthal directions.** Switching area in Hall bar sample with an azimuthal rotation of 0°**(a)**, 30°**(b)**, 60°**(c)**, 90°**(d)**, 120°**(e)**, 150°**(f)**, 180°**(g)**, 210°**(h)**, 240°**(i)**, 270°**(j)**, 300°**(k)**, 330°**(l)**. The injection direction of current pulse is marked at one port of the current channel.

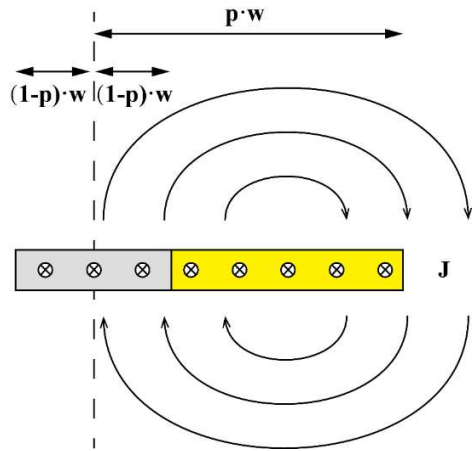

**Fig. S5 Schematic illustration of Oersted field generated by evenly distributed current in Hall bar.** The dashed line marks the calculation position. The grey region marks the region whose net contribution to the Oersted field is canceled out.

**Table S1. Main parameters for simulation**

| Parameter | Description | Default value |
|---|---|---|
| $M_S$ | Saturation magnetization | $1\times10^6$ A·m$^{-1}$ |
| $H_k$ | Anisotropy field | $1.6\times10^5$ A·m$^{-1}$ |
| $\alpha$ | Damping factor | 0.1 |
| $J_{RKKY}$ | Antiferromagnetic RKKY coupling energy density | 0.2 mJ·m$^{-2}$ |
| $D_{1,2}$ | Interlayer DMI energy density | 0.04 mJ·m$^{-2}$ |
| $t_{FM}$ | Ferromagnet layer thickness | 1 nm |
| $\theta_{SOT}$ | Spin Hall angle | ±0.3 |
| $H_{ext}$ | External field | 0 A·m$^{-1}$ |
| $t$ | Pulse duration | 1.5 ns |
| $dt$ | Step of time evolution in simulation | 0.01 ps |
| $\hbar$ | Reduced Planck constant | $1.054\times10^{34}$ J·s |
| $\sigma$ | Spin polarization | [0,1,0] |
| $\mu_0$ | Vacuum Permeability | $4\pi\times10^{-7}$ H·m$^{-1}$ |
| $\gamma$ | Gyromagnetic ratio | $1.76\times10^{11}$ rad·s$^{-1}$·T$^{-1}$ |
| $e$ | Elementary charge | $1.602\times10^{-19}$ C |